# Graph isomorphism testing boosted by path coloring

Thomas E. Portegys, Dialectek, *portegys@gmail.com*

## Abstract

A method for improving the efficiency of graph isomorphism testing is presented. The method uses the structure of the graph colored by vertex hash codes as a means of partitioning vertices into equivalence classes, which in turn reduces the combinatorial burden of isomorphism testing. Unrolling the graph into a tree at each vertex allows structurally different regular graphs to be discriminated, a capability that the color refinement algorithm cannot do.

## Introduction

Numerous uses can be found for unique and concise graph identifiers: graphs could then be counted, sorted, compared and verified more easily. For example, chemical compounds could be specified by identifying their constituent molecules represented by graphs of spatial and bonding relationships between atoms. However, a problem with developing a method for identifying graphs is that graphs are very general objects. Uniquely identifying vertices and edges solves the problem but begs the question, since the problem then becomes how to arrive at these identifiers in a uniform fashion (Sayers and Karp, 2004).

A method developed by Portegys (2008) identifies vertices by computing an MD5 hash (Rivest, 1992) for a tree of nodes rooted at each vertex. A vertex tree is composed by unrolling reachable vertices. Once each vertex is hashed, the vertex hashes are sorted and hashed to yield a hash for the graph, a technique similar to that used by Melnik and Dunham (2001) and Bhat (1980).

The vertex hashing can be seen as a coloring process, along the lines of the well-known *color refinement* algorithm (Arvind et al., 2015; Grohe et al., 2014). Given a graph $G$, the color refinement algorithm (or *naive vertex classification*) iteratively computes a sequence of colorings $C^i$ of $V(G)$. The initial coloring $C^0$ is uniform.
Then,

$$C^{i+1}(u) = \{\{ C^i(a) : a \in N(u) \}\}, \tag{1}$$

where $\{\{...\}\}$ is a multiset operator. Note that $C^1(u) = C^1(v)$ iff the two vertices have the same degree.

Thus the coloring begins with a uniform coloring of the vertices of the graph and refines it step by step so that, if two vertices have equal colors but differently colored neighborhoods (with the multiplicities of colors counted), then these vertices get new different colors in the next refinement step. The algorithm terminates as soon as no further refinement is possible.

Graphs are isomorphic if there is a consistent mapping between their vertices (Karp, 1972). For isomorphism testing of graphs *G* and *H*, the color refinement algorithm concludes that *G* and *H* are non-isomorphic if the multisets of colors occurring in these graphs are different. If this happens, the conclusion is correct. However, not all non-isomorphic graphs are distinguishable, or *amenable*, to color refinement. The simplest example is given by any two non-isomorphic regular graphs of the same degree with the same number of vertices, such as those shown in Figure 1.

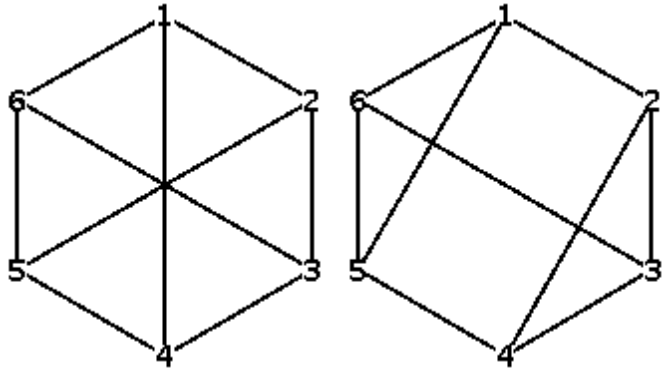

Figure 1 – Regular non-isomorphic graphs.

Graph isomorphism testing, a problem that has long been believed to be of non-polynomial complexity (NP), has recently been the subject of renewed attention (Babai, 2015). Graph isomorphism is also of practical use in a number of areas, including mathematical chemistry and electronic design automation. A number of isomorphism testing algorithms are in use, e.g. the Ullman (1976) and Schmidt-Druffel (1976) algorithms. The Ullmann algorithm is one of the most commonly used for graph isomorphism because of its generality and effectiveness (Cordella, et. al., 2001).

Graph coloring partitions vertices into equivalence classes of identical colors, which can reduce the complexity of isomorphism testing significantly. While the color refinement algorithm uses vertex degree as a shallow means of grouping vertices, using deep vertex hashing as a coloring method produces a finer discrimination of structure such that non-isomorphic regular graphs can be differentiated. For example, the hashes for the graphs depicted in Figure 1 will be different.

## Description

This section describes the hashing algorithm.

### *Graph format*

Using a pseudo-C++ notation, the following define a graph vertex and edge:

```
Vertex
{
    int label;
    Edge edges[];
};
```

```
Edge
{
    int label;
    Vertex source;
    Vertex target;
    bool directed;
};
```

This general scheme allows for a number of graph variations: labeled/unlabeled (using null labels), directed/undirected (for undirected, source and target are synonymous), and multigraphs.

## *Algorithm*

The following object is used to construct MD5 hash codes based on vertex graph neighborhoods:

```
VertexCoder
{
      Vertex vertex;
      vector<Vertex *> vertexBranch;
      unsigned char code[MD5_SIZE];
      void encode(bool hashLabels);
      void expand();
      void contract();
};
```

The algorithm iteratively expands each vertex in the graph into tree of coder objects representing the vertices and edges in its neighborhood. Branching terminates when a duplicate vertex appears in a branch, at which point the terminal coder takes on the hashed value of the distance of the first appearance of the vertex in the branch. The graph hash code is then constructed by sorting and hashing the vertex codes.

```
// Encode graph.
// The boolean argument allows labels to be included in the
// hash calculation.
void encode(bool hashLabels)
{
   if (vertex != NULL)
   {
      expand();
   }
   int numChildren = children.size();
   for (i = 0; i < numChildren; i++)
   {
      children[i]->coder->encode(hashLabels);
      children[i]->coder->contract();
   }
   sort(children);
   input = new unsigned char[HASH_INPUT_SIZE];
```

```
if (vertex != NULL)
{
   if (hashLabels)
   {
      append(input, vertex->label);
   }
   if (numChildren > 0)
   {
      for (i = 0; i < numChildren; i++)
      {
         edge = children[i]->edge;
         if (hashLabels)
         {
            append(input, edge->label);
         }
         if (edge->directed)
         {
            if (edge->source == vertex)
            {
               append(input, 1);
            }
            else
            {
               append(input, 0);
            }
         }
         else
         {
            append(input, 2);
         }
      }
   }
   else
   {
      for (i = 0; i < vertexBranch.size(); i++)
      {
         if (vertex == vertexBranch[i])
         {
            break;
         }
      }
      i++;
      append(input, i);
   }
}
for (i = 0; i < numChildren; i++)
{
   append(input, children[i]->coder->code);
}
code = MD5hash(input);
}
```

```
// Expand coder.
void expand()
{
   vector<Vertex *> childVertexBranch;
   Vertex *childVertex;
   VertexCoder *child;

   for (i = 0; i < vertexBranch.size(); i++)
   {
      if (vertex == vertexBranch[i])
      {
         return;
      }
      childVertexBranch.push_back(vertexBranch[i]);
   }
   childVertexBranch.push_back(vertex);
   for (i = 0; i < vertex->edges.size(); i++)
   {
      if (vertex == vertex->edges[i]->source)
      {
         childVertex = vertex->edges[i]->target;
      }
      else
      {
         childVertex = vertex->edges[i]->source;
      }
      child = new VertexCoder(childVertex, childVertexBranch);
      children.push_back(child);
   }
}

// Contract coder.
void contract()
{
   for (i = 0; i < children.size(); i++)
   {
      delete children[i]->coder;
      delete children[i];
   }
   children.clear();
}
```

Since each of $N$ vertices unrolls a tree of potentially $N$ nodes, the algorithm complexity is $O(N^2)$. A proof of the algorithm appears challenging, but is currently underway. These initial results are provided with the hope of eliciting further insights.

## *Example*

The method is illustrated through an example. Consider the simple directed graph shown in Figure 2. The vertices and edges are labeled for illustrative purposes, but the algorithm works for unlabeled vertices and edges as well as undirected edges.

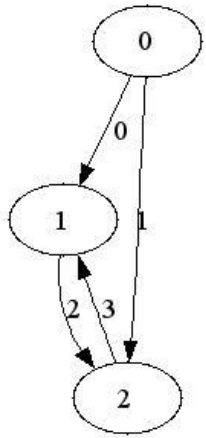
Figure 2 – A simple directed graph.

Before the first call to encode(), the coder is configured as in Figure 3. This configuration reveals nothing about the edges in the graph, and thus must always be expanded.

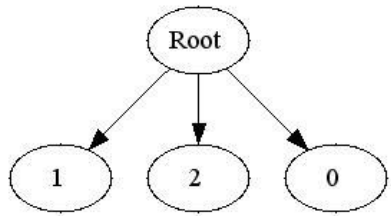
Figure 3 – Initial coder configuration.

After the first expansion, the coder appears as in Figure 4. The (f) and (b) notation on the edges represent a directed edge in the source graph in the forward and backward direction respectively. Note that for vertex 0, there are 2 forward edges to vertices 1 and 2. For vertex 1, there is a forward and backward edge to vertex 2, and a backward edge to vertex 0. Vertex 2 has a forward and backward edge to vertex 1, and a backward edge to vertex 0. Although Figure 4 shows expanded vertices concurrently, in actuality a vertex is removed through contraction after its hash values is obtained by its parent. The expansion continues until each branch reaches a duplicate vertex.

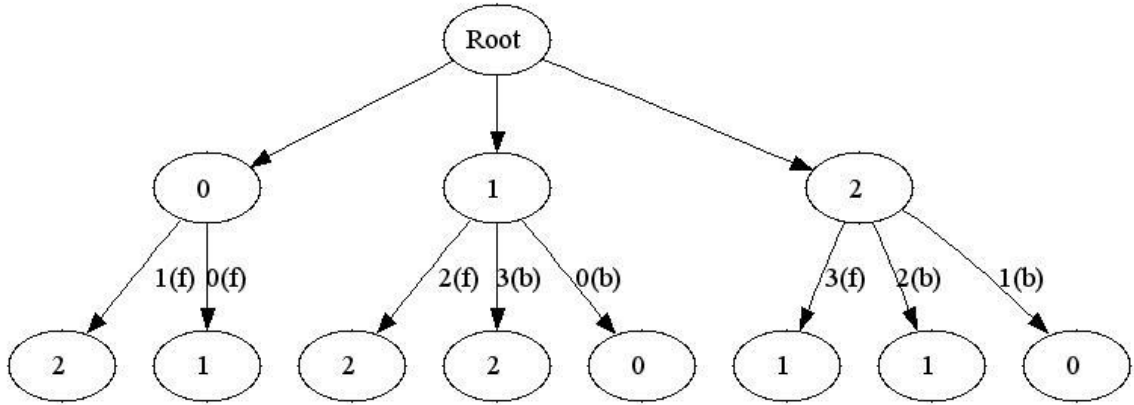

Figure 4 – First expansion.

Figure 5 depicts how the terminal coder values for two branches in the expansion tree are assigned. On the left branch, the terminal is assigned a value of 1, as it is a duplicate of the first coder (shown double-bordered), which 1 distant from the root. Likewise the right terminal is a duplicate of the second coder and is assigned a value of 2. If the terminal coder has no duplicate, it is assigned the length of the branch as a value.

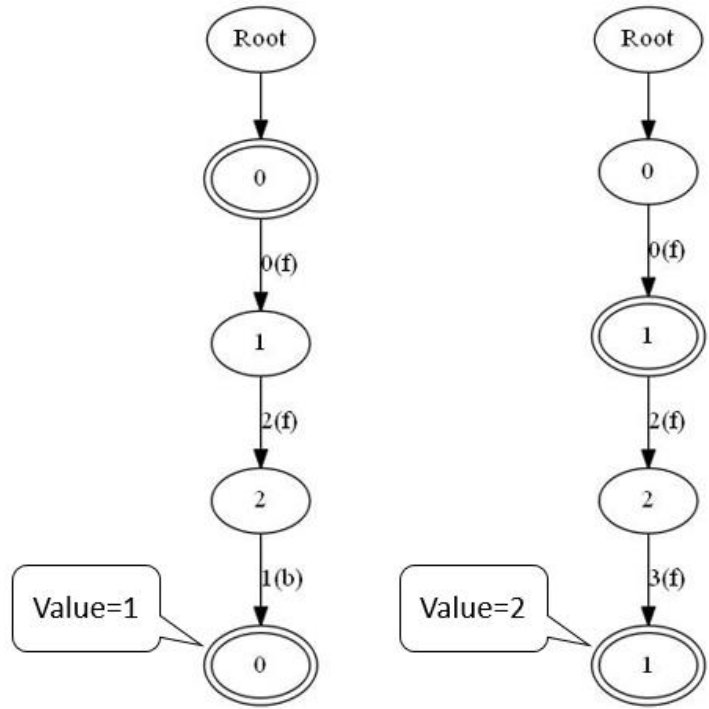

Figure 5 – Branch terminal coder values.

## Results

To highlight the potency of vertex partitioning using a coloring algorithm such as hashing, a comparison with brute force isomorphism testing is given in Table 1. For each isomorphism test, a graph with random edge connections and random vertex and edge labels is generated. Its isomorph is created by adding to each vertex and edge label the maximum label value of the original graph plus one. The number of search combinations to test isomorphism was measured. The relatively small graphs rapidly explodes in complexity for the brute force method, while the hashed method remains remarkably flat.

| Vertices x edges | Brute force | Hashed |
| --- | --- | --- |
| 5x5 | 14.1 | 15.5 |
| 5x10 | 16.2 | 26.9 |
| 10x10 | 5465.9 | 29.6 |
| 10x20 | 1593.8 | 43.2 |
| 15x15 | 5108975 | 44.2 |

Table 1 – Isomorphism testing comparison.

The results presented in Portegys (2008) validated the ability of the algorithm to uniquely hash unique graphs. Here we focused on the ability of the hash algorithm to discriminate regular graphs in comparison to the color refinement algorithm. A graph generation package (Johnsonbaugh and Kalin, 1991) was used to generate pairs of regular graphs with varying number of vertices and degrees. Each pair consisted of graphs having the equal quantities of vertices and equal degree. Some pairs were by chance isomorphic and others were non-isomorphic. The color refinement algorithm, as expected, classified all the pairs as isomorphic. The hash algorithm correctly distinguished all isomorphic and non-isomorphic pairs.

## Conclusion

A method for boosting the efficiency of graph isomorphism testing has been presented. The method is able to discriminate graphs that elude the color refinement algorithm. The method builds on a previously developed technique for identifying graphs using MD5 hashing.

The C++ code can be found here:
http://sourceforge.net/projects/graph-hashing/